\documentstyle[12pt,psfig,epsfig,aasms4,amsfonts]{article}
\def\degr{\hbox{$^\circ$}}

\def\arcsec{\hbox{$^{\prime\prime}$}}

\def\lsim{\mathrel{\hbox{\rlap{\lower.55ex \hbox {$\sim$}}\kern-.0em
\raise.4ex \hbox{$<$}}}} 
\def\gsim{\mathrel{\hbox{\rlap{\lower.55ex \hbox {$\sim$}}\kern-.0em
\raise.4ex \hbox{$>$}}}} 
\newcommand{\gpm}[3]{$#1^{+#2}_{-#3}$}

\def\grb{GRB\thinspace{991208}}

\begin{document}

\title{Continued Radio Monitoring of the Gamma Ray Burst 991208}

\author{
T. J. Galama\altaffilmark{1},
D. A. Frail\altaffilmark{2},      
R.    Sari\altaffilmark{3},   
E.    Berger\altaffilmark{1},     
G. B. Taylor\altaffilmark{2},
\&     
S. R. Kulkarni\altaffilmark{1}       
}

\vspace{-3mm}

\altaffiltext{1}{Division of Physics, Mathematics and Astronomy,
 California Institute of Technology, MS~105-24, Pasadena, CA 91125}

\altaffiltext{2}{National Radio Astronomy Observatory, P.O. Box 0,
Socorro, NM 87801}

\altaffiltext{3}{California Institute of Technology,
 Theoretical Astrophysics  103-33, Pasadena, CA 91125}


\begin{abstract}
  We present radio observations of the afterglow of the bright
  $\gamma$-ray burst \grb\ at frequencies of 1.4, 4.9 and 8.5 GHz,
  taken between two weeks and 300 days after the burst. The
  well-sampled radio light curve at 8.5 GHz shows that the peak flux
  density peaked about 10 days after the burst and decayed thereafter
  as a power-law F$_{\rm R}\propto t^{-1.07\pm0.09}$. This decay rate
  is more shallow than the optical afterglow of \grb\ with F$_{\rm
    o}\propto t^{-2.2}$, which was measured during the first week.
  These late-time data are combined with extensive optical, millimeter
  and centimeter measurements and fitted to the standard relativistic
  blast wave model. In agreement with previous findings, we find that
  an isotropic explosion in a constant density or wind-blown medium
  cannot explain these broadband data without modifying the assumption
  of a single power-law slope for the electron energy distribution.  A
  jet-like expansion provides a reasonable fit to the data. In this
  case, the flatter radio light curve compared to the optical may be
  due to emission from an underlying host galaxy, or due to the
  blastwave making a transition to non-relativistic expansion. The
  model that best represents the data is a free-form model in which it
  is assumed that the broadband emission originates from a synchrotron
  spectrum, while the time-evolution of the break frequencies and peak
  flux density are solved for explicitly. Although the decay indices
  for most of the synchrotron parameters are similar to the jet model,
  the evolution of the cooling break is unusually rapid $\nu_{\rm c}
  \propto t^{-2}$, and therefore requires some non-standard evolution
  in the shock.
\end{abstract} 




\section{Introduction}

The afterglow of \grb\ was one of the brightest observed to date.
Shortly after the coordinates of the $\gamma$-ray error box were made
known, a previously uncataloged radio source was discovered near its
center.  Based on the source location, its inverted spectrum and
compactness, it was a strong candidate for being the radio counterpart
of \grb\ (\cite{fra99}).  This was confirmed by the subsequent
discovery of an optical transient at the same location as the radio
source (\cite{cas+99}).  Follow-up optical observations taken from 2
to 7 days after the burst showed that the spectral and temporal
characteristics of this burst could be represented by power-law
indices $\alpha = 2.2\pm 0.1$, and $\beta = 0.75\pm 0.03$, where F$(t)
\propto t^{-\alpha} \nu^{-\beta}$ (\cite{smp+00}). With added data,
Castro-Tirado et al.~(2001)\nocite{csg+01} determined $\alpha = 2.3\pm
0.07$, and a somewhat steeper $\beta = 1.05\pm 0.09$. The $\gamma$-ray
properties, the localization and the subsequent afterglow discovery
for \grb\ are presented in more detail by Hurley et
al.~(2000)\nocite{hcm+00}.

\grb\ was also very bright at millimeter wavelengths (\cite{she+99})
and as a consequence, well-sampled spectra and light curves covering
frequencies between 1.43 to 240 GHz were taken during the initial two
weeks following the burst (\cite{gbb+00}; Paper I). These observations
made it possible, for the first time, to trace the time-evolution of
the synchrotron spectrum directly without resorting to specific models
for the evolution of the characteristic frequencies and the peak flux
density.  In comparing these results with model predictions it was
found that spherically symmetric explosions in homogeneous or
wind-blown circumburst media could be ruled out. A model in which the
relativistic outflow is collimated ({\em i.e.,} a jet) was shown to
account for the observed evolution of the synchrotron parameters, the
rapid decay at optical wavelengths, and the observed radio to optical
spectral flux distributions, provided that the jet transition has not
been fully completed in the first two weeks after the event.

This high quality dataset stimulated other modeling efforts for \grb.
Li and Chevalier (2001)\nocite{lc01} argued that an isotropic
explosion in a wind-blown ambient medium described the data equally
well. However, in order to account for the fast decay of the optical
light curves with respect to the radio light curves, they invoked a
non-standard break in the electron energy distribution. Panaitescu \&
Kumar (2002)\nocite{pk02} also required a broken power-law and they
concluded that the best fit to the data was obtained for a jet-like
outflow with an opening angle of about 15$^\circ$ expanding into
either a constant density or wind-blown circumburst medium. Dai \& Gou
(2001)\nocite{dg01} present a variation on these other efforts,
arguing that the steep optical decay is due to an anisotropic jet
viewed off axis, which is expanding into a wind from a red supergiant.

In view of the disparate conclusions reached on the basis of this
early afterglow data, there is some hope that with further monitoring
of the afterglow it may be possible to distinguish between these
models.  Accordingly, in this paper we present radio observations of
\grb\ made with the VLA and the Very Large Baseline Array (VLBA) which
began at the end of the two week period covered in Paper I and
continued until the burst faded below our detection limits. We focus
on the interpretation of the observations in terms of the predictions
made by relativistic blast wave models. The observations and results
are presented in \S\ref{sec:rea} and \S\ref{sec:results} and are
discussed in \S\ref{sec:dis}.

\section{Observations}\label{sec:rea}

\subsection{Very Large Array}

Observations at 1.43, 4.86 and 8.46 GHz were made using the NRAO Very
Large Array (VLA)\footnotemark\footnotetext{The NRAO is a facility of
  the National Science Foundation operated under cooperative agreement
  by Associated Universities, Inc.}. All observations were performed
in the default continuum mode in which, at each frequency, the full
100 MHz bandwidth was used in two adjacent 50-MHz bands. The
synthesized beam at 8.46 GHz was approximately 0.8\arcsec. The beam
size at other frequencies scales inversely with frequency. Gain
calibration was carried out by observing 3C\thinspace{48}. The array
phase was monitored by switching between the GRB and the phase
calibrators J1637+462 (at 8.46 GHz), J1658+476 (at 4.86 GHz) and
J1653+397 (at 1.43 GHz). Data calibration and imaging were carried out
with the {\em AIPS} software package following standard practice. We
performed phase self-calibration on the 1.43 GHz data. Flux densities
were determined by a least squares fitting of the Gaussian-shaped
synthesized beam to the source. For very faint detection levels ($< 3
\sigma$) we used the peak flux density at the position of the radio
counterpart (\cite{fra99}). A log of the observations is provided in
Table \ref{tab:obs}.

\subsection{Very Long Baseline Array}

Observations centered at 8.35 and 4.85 GHz were made using the NRAO
Very Long Baseline Array (VLBA) at four epochs. At both frequencies
the observations consisted of eight channels of bandwidth 8 MHz (64
MHz total bandwidth) spaced equally over a $\sim$ 0.5 GHz frequency
span.  The phases were calibrated using the calibrator source
1637+472. At each epoch the observations lasted $\sim$ 9 hours
(divided over the source, calibrators and frequency switches).  These
observations were designed for an interstellar scintillation
experiment (requiring substantial time and frequency span), the
results of which will be presented in a separate paper.  Here, in
Table \ref{tab:vlbaobs}, we report just the time-averaged flux
densities.

\section{Results}\label{sec:results}

Multiwavelength light curves of the event are presented in Figure
\ref{fig:obs}. Data are from this work (Tables \ref{tab:obs} and
\ref{tab:vlbaobs}) and Paper I. Optical observations are taken from
Castro-Tirado et al. (2001)\nocite{csg+01}, were we have only included
early data (less than 10 days after the event) to measure the `pure'
afterglow emission ({\em i.e.,} where the underlying host does not
contribute significantly).

The VLBA measurements may suffer from amplitude losses as a result of
imperfect phase referencing. The lightcurves indicate that such losses
may be significant since at 8.46 GHz the VLBA data seem to be
systematically lower than the VLA data.

The 8.46 and 4.86 GHz light curves rise to a maximum around 10 days.
While the 8.46 GHz light curve is smoothly rising to a maximum of
about 2 mJy during the first week, it undergoes a sudden drop to about
1 mJy (around 9 days after the burst) followed by a second peak
(around 12 days). Such erratic flux density variations are the
hallmark of interstellar scintillation (ISS). We will return to this
point in \S\ref{sec:iss}. The overall trend in Figure \ref{fig:obs} is
for the peak flux density to decline with decreasing frequency. A
linear least squares fit was carried out to the late time radio light
curve ($\Delta t$=53-293 days) at 8.46 GHz, and we derived a temporal
decay index $\alpha_{\rm R}=1.07\pm0.09$ ($\chi{_r}^2=1.07$, where
F$_{\rm R}\propto t^{-\alpha_{\rm R}}$).  This is to be compared with
the much steeper optical decay $\alpha_{\rm o}=2.2\pm0.1$ determined
at $\Delta t$=2-10 days (\cite{smp+00}).  Reconciling this difference
between the optical and radio decay indices is one of the main
challenges in modeling the afterglow of \grb\ (see \S\ref{sec:dis}).

\subsection{Interstellar Scintillation \label{sec:iss}}

The lines-of-sight towards most GRBs traverse a considerable
pathlength through the turbulent ionized gas of our Galaxy. As a
result, the radio emission from GRB afterglows is expected to be
affected by interstellar scattering (ISS). Indeed, as predicted by
Goodman (1997)\nocite{goo97}, ISS has been positively identified in
several cases (e.g. \cite{fkn+97}). The subject of ISS is a large one
(see review by Rickett 1990\nocite{ric90}), and encompasses a range of
observational phenomenology. Here we concern ourselves with estimating
the importance of ISS-induced modulation of our flux density
measurements as a function of time and frequency. For more detailed
treatments of ISS see Goodman (1997)\nocite{goo97} and Walker
(1998)\nocite{wal98}.

The character of the expected flux density fluctuations depend on
whether the scattering takes place in the ``strong'' or ``weak''
regimes. We define a transition frequency $\nu_0$ below which the
density irregularities can induce strong phase variations and above it
there is only weak scattering (WISS). Strong scattering can result in
both slow, broad-band changes, due to refractive interstellar
scintillation (RISS), and fast, narrow-band variations due to
diffractive interstellar scintillation (DISS). The value of $\nu_0$
along any given line-of-sight depends on the strength of the
scattering and the total pathlength. The amount of scattering is
parameterized by the scattering measure, SM, as modeled for the Galaxy
by Taylor and Cordes (1993)\nocite{tc93}. For convenience we assume
that the scattering along the line-of-sight takes place in a thin
screen located at $d_{scr}\sim(h_z/2)\,{\rm sin}\vert{b}\vert^{-1}$,
where $h_z$ ($\simeq{1}$ kpc) is the exponential scale height of the
ionized gas layer (\cite{rey89}).

Radio afterglow emission will be modulated by ISS provided that the
source size $\theta_s$ is close to one of the three characteristic
angular scales (1) $\theta_w=\theta_{F_0}(\nu_0/\nu)^{1/2}$, (2)
$\theta_d=\theta_{F_0}(\nu_0/\nu)^{-6/5}$, and (3)
$\theta_r=\theta_{F_0}(\nu_0/\nu)^{11/5}$, where $\theta_{F_0}$ is the
angle subtended by the first Fresnel zone at the scattering screen for
$\nu=\nu_0$, i.e.  $\theta_{F_0}=({c/2 \pi \nu_0 d_{scr}})^{1/2}$. The
modulation index $m_p$ (i.e. the r.m.s. fractional flux variation) is
given by:

\begin{equation}
m_p=
   \cases{
     (\nu_0/\nu)^{17/12}          & for $\nu>\nu_{0}$ and $\theta_s<\theta_w$ ~~~(WISS), \cr
     1                              & for $\nu<\nu_{0}$ and $\theta_s<\theta_d$ ~~~(DISS), \cr
     (\nu_0/\nu)^{-17/30}         & for $\nu<\nu_{0}$ and $\theta_s<\theta_r$ ~~~(RISS). \cr
         }
\end{equation}       

The expansion of the afterglow may cause $\theta_s$ to eventually
exceed one or more of the three characteristic angular scales that
define the different scattering regimes. For ease of comparison in
Appendix \ref{sec:ThetaModel} we compute $\theta_s$ for three
different interstellar medium (ISM) models (a constant density ISM
model, a wind model [WIND] and a jet model [JET]; See
\S\ref{sec:syn}). Then the modulation will begin to ``quench'' as
follows:

\begin{equation}
m_\nu = {m_p\times}
   \cases{
     (\theta_w/\theta_s)^{7/6}          & for $\nu>\nu_{0}$ and $\theta_s>\theta_w$ ~~~(WISS), \cr
     (\theta_d/\theta_s)                & for $\nu<\nu_{0}$ and $\theta_s>\theta_d$ ~~~(DISS), \cr
     (\theta_r/\theta_s)^{7/6}          & for $\nu<\nu_{0}$ and $\theta_s>\theta_r$ ~~~(RISS). \cr
         }
\end{equation}       

\begin{figure}[b!] 
\centerline{\psfig{figure=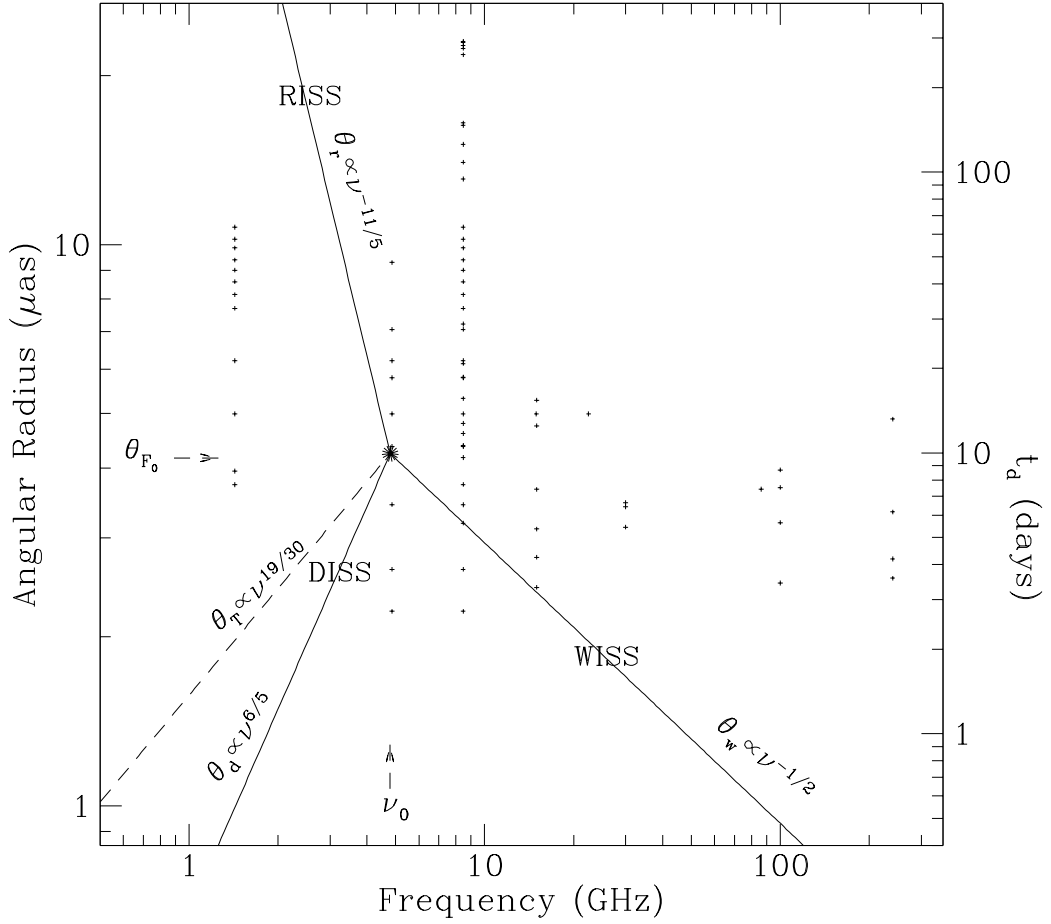,width=16cm}}
\caption{Interstellar scattering toward GRB\,991208 with the three
  regimes of scattering: weak (WISS), difractive (DISS) and refractive
  (RISS). The solid lines indicate the frequency dependence of each of
  three characteristic angular scales, joined at the transition
  frequency $\nu_0$ and the Fresnel scale $\theta_{F_0}$. On the right
  vertical axis we compute a model-dependent timescale, corresponding
  to the angular radius reached by a collimated fireball given on the
  left vertical axis.  The small crosses indicate the time and
  frequency of the measurements in Tables \ref{tab:obs} and
  \ref{tab:vlbaobs}.  See \S\ref{sec:results} for more details.}
\label{fig:iss}
\end{figure}

For GRB\,991208 at ($l, b)$ = (72\degr4,+42\degr6) we estimate SM =
2.26$\times 10^{-4}$kpc m$^{-20/3}$ from the Taylor and Cordes (1993)
model. Taking a typical scattering screen distance of $d_{scr}$ = 0.74
kpc we compute $\nu_0$ = 4.8 GHz and $\theta_{F_0} = 4.2$ $\mu$as
(Walker 1998). The difficulties in obtaining accurate estimates of SM
and $d_{scr}$ likely limit the accuracy of the $\nu_0$ and
$\theta_{F_0}$ values to $\pm$ 30\%. In Figure \ref{fig:iss} we plot
these values along with lines showing the different ISS regimes. On
the right hand side of the figure we compute a model-dependent
timescale for the fireball to reach the angular size given on the left
hand side. We have assumed that the emission from GRB\,991208
originates from a jet-like outflow for $t_{jet}\leq{2}$ days and used
the relevant expression for $\theta_s$ in Appendix
\ref{sec:ThetaModel}. At a redshift $z= 0.7055$ (\cite{das+99,ddb+99})
the angular distance $D_A=D_L/(1+z)^2$=0.45$\times{10}^{28}$ cm, where
$D_L$ is the luminosity distance. The source size is
$\theta_s=5.2(E_{52}/n_1)^{1/8}(t_d/15)^{1/2}$ $\mu$as, where $E_{52}$
is the energy in the blast wave in units of 10$^{52}$ erg and $n_1$ is
the density of the ambient medium in particles cm$^{-3}$. Thus the
observations in Tables \ref{tab:obs} and \ref{tab:vlbaobs} (and
indicated as small crosses in Figure \ref{fig:iss}) lie on either side
of $\nu_0$ and were taken during a time when $\theta_s$ expanded to
eventually exceed $\theta_{F_0}$.

It is apparent in Figure \ref{fig:iss} that most of the observations
were taken in the weak scattering regime ($\nu > \nu_0$).
Furthermore, for $\nu>15$ GHz, $\theta_s > \theta_w$ and therefore we
expect that the observed variations will be dominated by source
evolution and measurement error, not ISS. ISS starts to become
comparable to the measurement error at 15 GHz where it is
straightforward to show that $m_{15}=0.08(t_d/15)^{-7/12}$ (for
$\theta_s\geq\theta_w$).

Accounting for ISS is particularly important at 8.46 GHz since the
measurement error at this frequency is small ($<$ 5\%) while we expect
$m_{8.46}$ = 0.45 for $\theta_s<\theta_w$ and for $t_d\gsim{5}$ days
$m_{8.46}=0.24(t_d/15)^{-7/12}$. Since the 8.46 GHz light curve is so
well sampled, we can independently check on this prediction by
subtracting out any linear trend and computing the scatter. For
different assumed curves we find within the first two weeks of
observations an r.m.s. scatter of 20-30\%, suggesting that the
strength of scattering along this line-of-sight may be slightly weaker
than we have predicted.

We also expect large variations in the flux density at $\nu$ = 4.86
GHz but these are difficult to quantify, owning to the proximity of
$\nu_0$ (Walker 1998). For the purposes of model fitting we adopt
$m_{4.86}=0.5$ and for $t_d\gsim{10}$ days
$m_{4.86}=0.40(t_d/15)^{-7/12}$. 

The line $\theta_T = \theta_{F_0}(\nu_0/\nu)^{-19/30}$ indicates the
transition diameter that separates the region where the modulations
are RISS dominated (above $\theta_T$) from that were they are DISS
dominated (below $\theta_T$).  The observations at $\nu$ = 1.43 GHz
are clearly in the strong regime ($\nu < \nu_0$) where $\theta_d <
\theta_s < \theta_r$, and where RISS is expected to be significant with
$m_p$ = 0.5 on refractive timescales of a day or more.  Shorter
timescale ($t_{\rm DISS}\sim$3 hrs) variations due to DISS are also
expected but at a reduced level of 10-15\%, suppressed by the finite
source size (i.e. $\theta_s >> \theta_d$) and bandwidth averaging.
DISS, unlike WISS and RISS, is a {\it narrow-band} phenomenon with a
decorrelation bandwidth $\Delta\nu_d=\nu(\nu/\nu_0)^{17/5} < 100$ MHz,
the bandwidth of our observations.

\subsection{Synchrotron emission}
\label{sec:syn}

The radio/mm/optical to X-ray afterglow emission is believed to arise
from the forward shock of a relativistic blast wave that propagates
into the circumburst medium (see \cite{pir00,vkw00} for reviews). The
electrons are assumed to be accelerated to a power-law distribution in
energy, d$N/$d$\gamma_{\rm e} \propto \gamma_{\rm e}^{-p}$, for
$\gamma > \gamma_{\rm m}$, and radiate synchrotron emission; where
$\gamma_{\rm e}$ is the Lorentz factor of the electrons, $\gamma_{\rm
  m}$ is the minimum Lorentz factor, and $p$ the power-law index of
the electron energy distribution. Detailed calculations by Granot,
Piran and Sari (2000a,b)\nocite{gps99a,gps99b} and Granot and Sari
(2002)\nocite{gs02} have shown that to good approximation (to within a
few percent) the synchrotron afterglow spectrum can be described by
the following function

\begin{eqnarray}
   \label{eq:func}
   F(\nu;F_{\rm m},\nu_{\rm a},\nu_{\rm m},p) & = & F_{\rm m}\,(2)^{-1/n}[1-e^{-(\frac{\nu}{\nu_{\rm a}})^{-5/3}}]\left(\frac{\nu}{\nu_{\rm a}}\right)^{5/3}\,\nonumber\\
& \times &  \left[\left(\frac{\nu}{\nu_{\rm m}}\right)^{n/3} + 
\left(\frac{\nu}{\nu_{\rm m}}\right)^{-(p-1)n/2}\right]^{1/n}\,
\end{eqnarray}

Where, $F_{\rm m}$ is the synchrotron peak flux density at the peak
frequency $\nu_{\rm m}$, and $\nu_{\rm a}$ is the synchrotron
self-absorption frequency.  The index $n= -0.837$ (\cite{gs02}) and
controls the sharpness of the peak at $\nu_{\rm m}$. We have assumed
slow cooling, i.e.  $\nu_{\rm m} \ll \nu_{\rm c}$, where $\nu_{\rm c}$
is the synchrotron-cooling frequency. Castro-Tirado et
al.~(2001)\nocite{csg+01} have measured a steeper optical slope
$\beta$ than Sagar et al.~(2000)\nocite{smp+00} and have argued that
this is evidence that the afterglow observations of \grb\ require the
synchrotron cooling break to be located between radio and optical
wavelengths. We model a possible synchrotron cooling break by a sharp
spectral break with decrement $\Delta \beta$ = 0.5 at $\nu = \nu_c$,
and parameterization $\nu_c = C_c t^{-\alpha_c}$.

The evolution of the break frequencies $\nu_{\rm a}$ and $\nu_{\rm
  m}$, $\nu_{\rm c}$ and the peak flux density $F_{\rm m}$ follows
from assumptions on the dynamics of the relativistic blast wave. It is
common practice to assume three cases: i) propagation into a constant
density ambient medium (ISM; Sari, Piran and Narayan
1998\nocite{spn98}), ii) propagation into a wind-stratified ambient
medium (WIND; Chevalier and Li 1999, 2000\nocite{cl99,cl00}), and iii)
a blast-wave that is collimated in a jet (JET; Sari, Piran and Halpern
1999\nocite{sph99}; note that the evolution in the case of a jet is
rather insensitive to the density structure of the ambient medium; for
details see Kumar and Panaitescu 2000\nocite{kp00}). The three cases
have in common that the synchrotron parameters all evolve as power
laws in time ({\em i.e.,} $F_{\rm m} = C_F t^{-\alpha_F}$, $\nu_{\rm
  a} = C_{\rm a} t^{-\alpha_{\rm a}}$, $\nu_{\rm m} = C_{\rm m}
t^{-\alpha_{\rm m}}$ and $\nu_{\rm c} = C_{\rm c} t^{-\alpha_{\rm
    c}}$) but differ in the values of their decay indices and
coefficients. In addition to assuming model-dependent decay indices,
in this paper and in Paper I we solved for $\alpha_F$, $\alpha_{\rm
  a}$, $\alpha_{\rm m}$ and $\alpha_{\rm c}$ explicitly by model
fitting the above synchrotron spectrum (FREE). The advantage of this
approach is that it assumes only that the afterglow is due to
synchrotron emission with a power-law distribution of electron
energies, and it imposes no physical constraints on the geometry of
the outflow, the microphysics of the shock or the density structure of
the circumburst medium.

\subsection{Modeling the light curves \label{sec:modeling}}

We fit the function $F(\nu;C_F,C_{\rm a},C_{\rm m},C_{\rm c},
\alpha_F,\alpha_{\rm a},\alpha_{\rm m},\alpha_{\rm c},p)$ in
Eqn.~\ref{eq:func} to the observations presented in Paper I, this work
(Tables \ref{tab:obs} and \ref{tab:vlbaobs}) and the early time
optical data of Castro-Tirado et al.~(2001)\nocite{csg+01}. We
minimize the $\chi^2$ of the fit by the Levenberg-Marquardt
least-squares minimization scheme. We account for the effect of
interstellar scintillation by adding in quadrature the predicted
scatter due to ISS (see previous discussion) to the measurement error.
We determine the 68.3 \% (1$\sigma$) uncertainties in the parameters
by plotting contours in chisquared where $\chi^2$ is raised by 2.3 as
a function of the two relevant parameters and by leaving the remaining
parameters free. The result of the best fits are shown in Figure
\ref{fig:obs} and presented in Table \ref{tab:fit}.

In addition to this free-form model (FREE), we also fixed the
decay-rate parameters ($\alpha_F, \alpha_{\rm a}$, $\alpha_{\rm m}$,
and $\alpha_{\rm c}$) at the predicted values for the constant density
ISM, the WIND and the JET model and repeated the fit. The results of
the JET fit are shown in Figure \ref{fig:obsj} and presented in Table
\ref{tab:fit}. As expected from the earlier attempts in Paper I, the
standard WIND and constant density ISM fits were rejected as
unsuitable ($\chi_{r}^2>8$).  These fits do not account well for the
combination of the rapid decay at optical wavelengths and the more
shallow decay at radio wavelengths.

\section{Discussion}\label{sec:dis}

\begin{figure}[b!] 
\centerline{\psfig{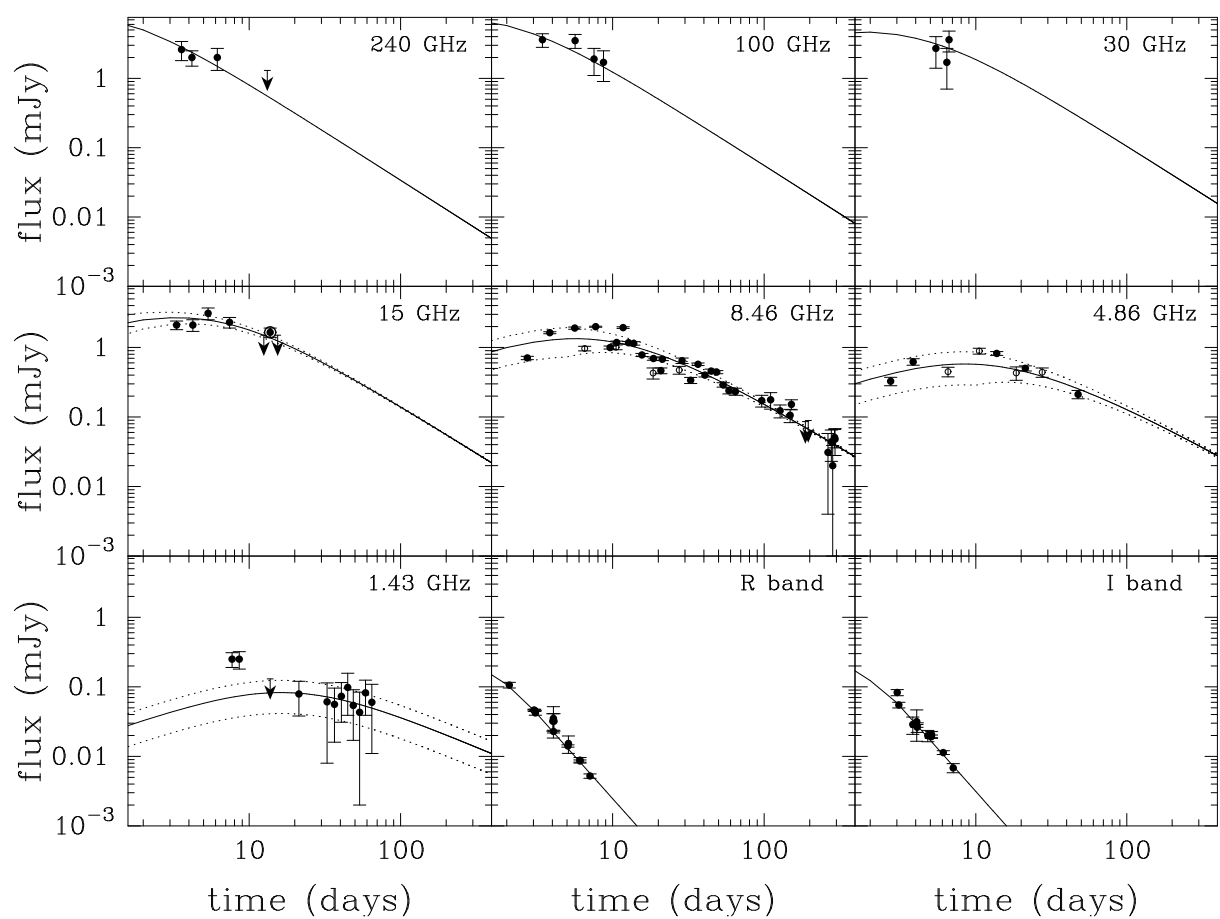}}
\caption{The radio to optical lightcurves of GRB\,991208.  Data are
  from this work (Table \ref{tab:obs}) and Paper I. Optical
  observations are from Castro-Tirado {\it et al.} (2001). Open
  symbols indicate the VLBA data (Table \ref{tab:vlbaobs}). The solid
  line indicates the best free-form fit to the light curves and the
  dotted lines are the predicted r.m.s. scatter due to interstellar
  scintillation. See \S\ref{sec:results} for more details.}
\label{fig:obs}
\end{figure}

The principal challenge in modeling the afterglow of \grb\ is in
simultaneously fitting for the decay of the radio ($\alpha_{\rm
  R}=1.07\pm0.09$) and optical ($\alpha_{\rm o}=2.2\pm0.1$) light
curves.  The model that best represents the data is the free-form
model (FREE) in which all the parameters in Eqn.~\ref{eq:func} were
solved for.  The FREE model shown in Figure \ref{fig:obs} and
summarized in Table \ref{tab:fit} differs from the original model
presented in Paper I in several respects. Most notably, with the
addition of the optical data from Castro-Tirado et
al.~(2001)\nocite{csg+01}, it has been necessary to allow for a
cooling break between the optical and millimeter bands
(\S\ref{sec:syn}).  This has the effect of flattening the value of $p$
which improves the fit to the decay of the optical light curve. A
cooling break in the spectrum also alters the temporal decline of the
synchrotron parameters but the values in Table \ref{tab:fit} agree,
within the errors, to those derived in Paper I.  The FREE model
resolves the differences between $\alpha_{\rm R}$ and $\alpha_{\rm o}$
by allowing for a relatively shallow decay of the peak flux density,
{\em i.e.} F$_{\rm m} \propto t^{-0.34}$, and an unusually rapid
evolution of the cooling frequency to lower frequencies, {\em i.e.}
$\nu_{\rm c} \propto t^{-2.1}$. This is generally much faster than any
model yet proposed for GRB afterglows. For $\nu_{\rm c} \propto
t^{-2}$ this implies $\Gamma\times B^3$=constant (where $\Gamma$ is
the bulk Lorentz factor of the flow, and $B$ is the post-shock
magnetic field).  For any reasonable model both $\Gamma$ and $B$ are
{\it decreasing} with time unless either the circumburst density or
the fraction of shock energy in the B-field $\epsilon_B$ are
increasing in time. In a future paper (S. Yost, {\em in preparation})
we plan to explore such modifications to the standard model for a
number of well-studied afterglows.

Both Li and Chevalier (2001)\nocite{lc01} and Panaitescu \& Kumar
(2002)\nocite{pk02} suggested a revision to the basic afterglow model
in order to overcome the difficulties with jointly fitting the radio
and optical data for this burst. They proposed the existence of a
two-component electron energy distribution with different slopes above
and below some break frequency $\nu_b$. Reasonable fits were obtained
for either WIND and JET models provided that the radio (optical)
emission was dominated by the low (high) energy electrons. Li \&
Chevalier (2001) also made predictions for the future evolution of the
afterglow beyond the first two weeks, allowing further tests of their
model. The measured $\alpha_{\rm R}\sim 1.1$ for the 8.46 GHz light
curve is in good agreement with their predicted decay $\alpha_{\rm
  R}=1.25$.  The largest source of disagreement is in the low
frequency behavior.  In their WIND model the peak flux density at 1.43
GHz is predicted to be a factor of 10 times larger than what is
measured.  This is likely due to the relatively slow evolution of
$\nu_{\rm a}$ where $\alpha_{\rm a}=0.29$, a value that is closer to
the JET model ($\alpha_a=0.2$) than the WIND model ($\alpha_a=0.6$).

\begin{figure}[b!] 
\centerline{\psfig{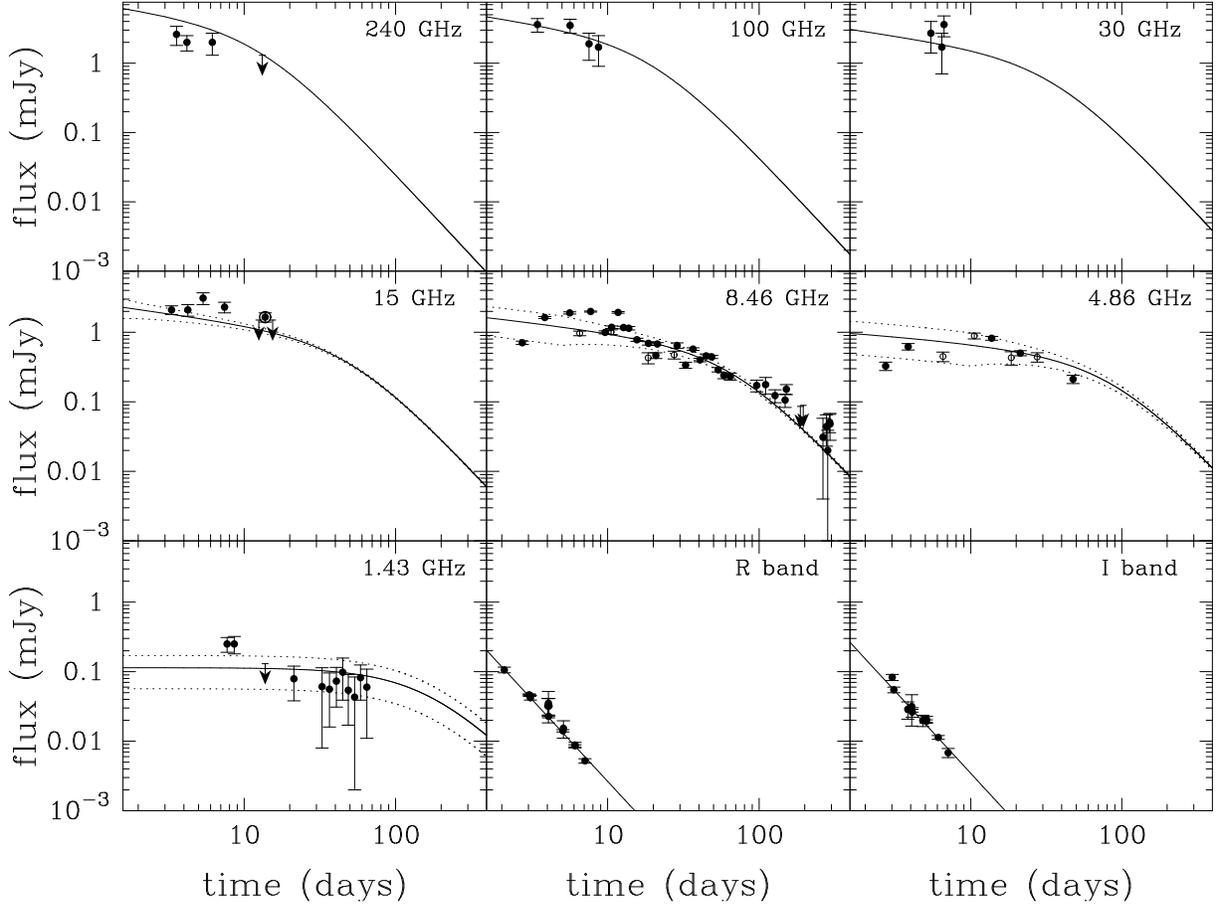}}
\caption{The radio to optical lightcurves of GRB\,991208. The solid
line indicates the best fit to the light curves for the JET
model and the dotted lines are the predicted r.m.s. scatter due to
interstellar scintillation. See \S\ref{sec:results} for more details.}
\label{fig:obsj}
\end{figure}

In Paper I when the spectral fits were compared with the expectations
for specific models, satisfactory agreement was found for the JET
model.  Nevertheless, there were a number of discrepancies, notably
the decline in the peak flux density F$_{\rm m}$ was too shallow, and
the value for the slope of the electron energy distribution $p=2.52$
was steeper than that implied by the decay of the optical light curves
$p\sim 2.2$. It was proposed that these problems with the JET model
could be reconciled if the jet was in transition to its fully
asymptotic behavior. In our new JET model (see Fig.~\ref{fig:obsj} and
Table \ref{tab:fit}) the addition of a cooling break improves the
overall quality of the fit considerably. Despite this agreement, there
is a suggestion in Fig.~\ref{fig:obsj} that the JET model is having
some difficulties with the late time radio measurements. Most of the
excess $\chi^2$ in the JET model results from the fact that the
late-time light curve ($\Delta t>$100 days) at 8.46 GHz is flat,
$\alpha_{\rm R}\sim 1.1$, whereas the JET model predicts that the flux
density will fall as F$_{\rm \nu} \propto t^{-p}$, in agreement with
$\alpha_{\rm o}\sim 2.2$.

There are two additional effects, not accounted for in our analysis,
that could make $\alpha_{\rm R}<\alpha_{\rm o}$. In the first
instance, \grb\ could have occurred in a radio-bright host galaxy
({\em e.g.}, Berger, Kulkarni \& Frail 2001\nocite{bkf01}).  Sokolov
et al.~(2001)\nocite{sfc+01} have used various optical methods to
estimate the star formation rate (SFR) for the host galaxy of \grb\ 
and derive an extinction-corrected SFR$\gsim 100$ M$_\odot$ yr$^{-1}$.
Although there are large uncertainties in such estimates, at $z=0.7$
such elevated SFR's should be detectable at centimeter wavelengths.
Assuming a standard starburst spectral energy distribution
(\cite{yc02}) we predict $F_\nu(1.43\,{\rm GHz})\approx 110$ $\mu$Jy,
significantly brighter than our measurements.  We can therefore infer
that the SFR in the host of \grb\ is probably no higher than a few
tens of M$_\odot$ yr$^{-1}$. Deeper radio observations taken a year or
more after the burst, when the emission from the afterglow is
negligible, are needed to provide a more accurate estimate of the star
formation rate. Another possible source of the flattening of the radio
light curves could be the transition of the relativistic blast wave to
non-relativistic expansion. Frail, Waxman, \& Kulkarni
(2000)\nocite{fwk00} modeled the decay of the radio light curves for
GRB\,970508 at $\Delta t\gsim$ 100 days as a non-relativistic
expansion. The expected decay rate in this phase of the evolution is
$\alpha_{\rm NR}=3(p-1)/2-3/5\approx 1.2$ for $p=2.2$.  Thus, it is
possible that the difference in decay rates is due to a transition to
non-relativisitic expansion at $t\sim 40$ days. More sophisticated
modeling is needed which explicitly includes the evolving dynamics of
the blastwave ({\em e.g.,} \cite{hys+01}, \cite{pk02}) in order to
test this hypothesis.

In summary, with continued monitoring of the bright afterglow of \grb,
it has been possible to track the evolution of its radio light curve
out to nearly 300 days after the burst. The power-law decline of the
optical light curve at early times is considerably steeper than the
radio light curve at late times. There are several different
explanations for reconciling this disparate behavior, including
modifying the standard relativistic model with non-standard electron
distributions, and with circumburst densities or equipartition
parameters $\epsilon_B$ that increase with time. However, the
simplest explanation which is consistent with the data and requires
no significant modifications, is that the blastwave of \grb\ entered a
non-relativistic expansion phase several months after the burst.



\clearpage

\newpage
\scriptsize 
\begin{table}[t]
\begin{tabular}{lcccccccc}
\tableline
\tableline
Date & t$_{\rm start}$ - t$_{\rm end}$ & t$_{\rm int}$ &
$\Delta t$ & F$_{\rm 8.46}$ & F$_{\rm 4.86}$ & F$_{\rm 1.43}$ \\ 
(UT) & (hh:mm:ss) & (hh:mm:ss) & (days) & ($\mu$Jy) &
($\mu$Jy) & ($\mu$Jy) \\
\tableline
1999 Dec 23.80  & 19:03:30 - 19:17:30 & 11:20 & 15.61 & 782 $\pm$ 40    \\
1999 Dec 26.78  & 18:25:40 - 18:47:20 & 16:20 & 18.59 & 693 $\pm$ 38    \\
1999 Dec 28.98  & 23:14:50 - 23:38:10 & 35:30 & 20.79 & 464 $\pm$ 44    \\
1999 Dec 29.49  & 11:17:10 - 12:11:50 & 22:00 & 21.30 & 679 $\pm$ 30    \\
1999 Dec 29.49  & 11:17:10 - 12:11:50 & 18:40 & 21.30 & & 501 $\pm$ 42 \\
1999 Dec 29.49  & 11:29:20 - 12:41:17 & 36:20 & 21.30 & & & 79 $\pm$ 41 \\
2000 Jan 5.94   & 22:25:00 - 22:37:10 &  9:30 & 28.75 & 646 $\pm$ 58 \\
2000 Jan 9.89   & 21:04:50 - 21:48:00 & 19:00 & 32.70 & 337 $\pm$ 34 \\
2000 Jan 9.89   & 20:58:00 - 21:37:50 & 14:25 & 32.70 & & & 61 $\pm$ 53 \\
2000 Jan 13.79  & 18:13:10 - 19:33:00 & 31:30 & 36.60 & 574  $\pm$ 25 \\
2000 Jan 13.79  & 18:04:55 - 19:41:05 & 30:35 & 36.60 & & & 56 $\pm$ 40 \\    
2000 Jan 17.85  & 19:33:40 - 21:17:30 & 43:30 & 40.66 & 400 $\pm$ 23 \\
2000 Jan 17.85  & 19:29:40 - 21:00:10 & 26:30 & 40.66 & & & 73 $\pm$ 42 \\
2000 Jan 21.92  & 21:17:30 - 22:36:20 & 31:40 & 44.73 & 458 $\pm$ 31 \\
2000 Jan 21.92  & 21:11:50 - 22:43:50 & 28:30 & 44.73 & & & 98 $\pm$ 59 \\
2000 Jan 24.81  & 19:03:25 - 19:52:00 & 36:50 & 47.62 & & 212 $\pm$ 29 \\
2000 Jan 25.85  & 18:32:10 - 22:25:20 & 1:00:10 & 48.66 & 442 $\pm$ 23 \\
2000 Jan 25.85  & 18:28:00 - 22:19:40 & 1:02:55 & 48.66 & & & 54 $\pm$ 37 \\
2000 Jan 30.85  & 18:42:30 - 22:22:30 & 1:20:50 & 53.66 & 290 $\pm$ 21 \\
2000 Jan 30.85  & 18:38:10 - 22:13:20 & 1:02:55 & 53.66 & & & 43 $\pm$ 41 \\  
2000 Feb 4.80   & 18:22:50 - 19:59:20 & 38:00 & 58.61 & 239 $\pm$ 24 \\
2000 Feb 4.80   & 18:18:30 - 20:06:30 & 31:45 & 58.61 & & & 82 $\pm$ 43 \\
2000 Feb 10.78  & 17:59:10 - 19:18:00 & 31:30 & 64.59 & 232 $\pm$ 27 \\
2000 Feb 10.78  & 17:55:00 - 19:26:40 & 26:55 & 64.59 & & & 60 $\pm$ 49 \\
2000 Mar 13.63  & 14:42:50 - 15:37:20 & 40:50 & 96.44 & 172 $\pm$ 33 \\
2000 Mar 27.57  & 13:34:20 - 13:48:50 & 11:50 & 110.38 & 177 $\pm$ 47 \\
2000 Apr 13.67  & 15:43:50 - 16:36:20 & 38:50 & 127.48 & 123 $\pm$ 26 \\
2000 May 4.62   & 14:27:40 - 16:20:50 & 2:01:10 & 148.43 & 106 $\pm$ 23 \\
2000 May 7.64   & 14:30:50 - 16:24:50 & 1:24:20 & 151.45 & 152 $\pm$ 25 \\
2000 Jun 6.45   & 10:31:10 - 11:09:30 & 24:40 & 181.26 & 38 $\pm$ 40 \\ 
2000 Jun 14.40  & 09:24:10 - 10:01:10 & 27:00 & 189.21 & 29 $\pm$ 30 \\  
2000 Aug 27.76  & 22:15:30 - 23:08:10 & 42:00 & 263.57 & 31 $\pm$ 27     \\
2000 Sep 10.85 & 20:28:10 - 21:19:10 & 01:25:20 & 277.66 & 44 $\pm$ 21 \\
2000 Sep 16.78 & 18:37:00 - 21:25:30 & 02:28:50 & 283.59 & 20 $\pm$ 19 \\
2000 Sep 24.77 & 18:35:30 - 20:53:40 & 02:02:03 & 291.58 & 51 $\pm$ 15 \\
2000 Sep 27.10 & 02:19:10 - 03:40:30 & 01:05:30 & 293.91 & 48 $\pm$ 20 \\

\tableline
\tableline
\end{tabular} 
\caption[c]{Summary of VLA observations of GRB\,991208. The date,
  start time $t_{\rm start}$ and end time $t_{\rm end}$ of the
  observations are given in UT. Also, the total time on source t$_{\rm
    int}$, the time passed since the event $\Delta t$ and the fluxes
  at 8.46, 4.86 and 1.43 GHz are given. Observations taken before Dec
  22 1999 can be found in Paper I. \label{tab:obs}}
\end{table} 
\normalsize
\clearpage

\newpage
\scriptsize 
\begin{table}[t]
\begin{tabular}{lcccccccc}
\tableline
\tableline
Date & 
$\Delta t$ & F$_{\rm 8.35}$ & F$_{\rm 4.85}$ \\ 
(UT) & (days) & ($\mu$Jy) &
($\mu$Jy)  \\
\tableline
1999 Dec 14.73  & 6.54  & 967 $\pm$ 55  & 448 $\pm$ 67  \\
1999 Dec 18.72  & 10.53 & 1008 $\pm$ 60 & 888 $\pm$ 74  \\
1999 Dec 26.70  & 18.51 & 430 $\pm$ 75  & 431 $\pm$ 92  \\
1999 Jan  4.69  & 27.50 & 474 $\pm$ 56  & 440 $\pm$ 64 \\
\tableline
\tableline
\end{tabular} 
\caption[c]{Summary of VLBA observations of GRB\,991208: the date, the
time passed since the event $\Delta t$ and the fluxes at 8.35, and
4.85 GHz.}\label{tab:vlbaobs}
\scriptsize
\end{table} 
\normalsize

\footnotesize
\begin{deluxetable}{llllllllllll}
  \tabcolsep0in\footnotesize \tablewidth{\hsize} \tablecaption{Best
    Fit Model Parameters for \grb\,: the value of $p$, the peak flux
    density F$_{\rm m} = C_F \times t_{\rm d}^{-\alpha_F}$, the
    synchrotron self-absorption frequency $\nu_{\rm a} = C_{\rm a}
    \times t_{\rm d}^{-\alpha_{\rm a}}$, the peak frequency $\nu_{\rm
      m} = C_{\rm m} \times t_{\rm d}^{-\alpha_{\rm m}}$, the cooling
    frequency $\nu_{\rm c} = C_{\rm c} \times t_{\rm d}^{-\alpha_{\rm
        c}}$, the reduced chisquared $\chi_{\rm r}^2$ and the degrees
    of freedom (d.o.f.) of the fit.
\label{tab:fit}}
\tablehead { 
\colhead {p} & 
\colhead {$C_F$} 
& \colhead {$C_{\rm a}$}
& \colhead {$C_{\rm m}$} 
& \colhead {$C_{\rm c}$} 
& \colhead {$\alpha_F$} 
& \colhead {$\alpha_{\rm a}$} 
& \colhead {$\alpha_{\rm m}$} 
& \colhead {$\alpha_{\rm c}$} 
& \colhead {$\chi_{\rm r}^2$} 
& \colhead {d.o.f} 
& \colhead {Note} \\ 
& \colhead {(mJy)} 
& \colhead {(GHz)} 
& \colhead {(THz)} 
& \colhead {(THz)} 
& \\ } \startdata

\gpm{2.14}{0.16}{0.09} & \gpm{7.7}{15.7}{2.3} & \gpm{28}{72}{12} &
\gpm{0.34}{0.29}{0.30} & \gpm{2670}{1200}{600} & 
\gpm{0.34}{0.28}{0.74} & \gpm{0.29}{0.21}{0.17} &
\gpm{1.9}{1.8}{1.2} & \gpm{2.08}{0.20}{0.11} & 
1.52 & 104 & FREE \\

\gpm{2.36}{0.19}{0.03} & \gpm{17.8}{4.2}{1.2} & \gpm{6.3}{2.0}{1.9}&
\gpm{39.4}{60.1}{3.10} &  \gpm{3.0}{1.8}{2.1} & 
1.0 & 0.2 & 2.0 & 0.0 & 
1.89 & 108 & JET \\

\enddata \tablecomments{Fixed parameters can be recognized by the fact
that they have no quoted uncertainties.}
\end{deluxetable} 
\normalsize

\newpage
\clearpage

\appendix

\section{Observed Angular Size \label{sec:ThetaModel}}
The relation between observed time, $T$, real time $t$, the radius $R$ and
the angle $\mu =\cos \theta $ is given by

\begin{equation}
T=t-\,\mu R/c.  \label{Trtmu}
\end{equation}
$R$ and $t$ are related by some hydrodynamic solution given by the
dynamics of the shock, so that in a given observed time $T$, $R$ is a
function of $\mu$ only. Shock dynamics are usually described by a
powerlaw, $\gamma \sim R^{-m/2}$. Therefore,
\[
t=\frac{R}{c}\left[ 1+\frac{1}{4(m+1)\gamma ^{2}}\right] .
\]
Substituting into equation \ref{Trtmu} gives 
\[
cT/R=1-\mu +\frac{1}{4(m+1)\gamma ^{2}}.
\]
Following Sari (1998)\nocite{sar98} we define $\gamma _{L}$ to be the
Lorentz factor of the fluid behind the shock on the line of sight and
$R_{L}$ to be the shock radius on the line of sight, both at time
$T$. We get
\begin{equation}
cT=\frac{R_{L}}{4(m+1)\gamma _{L}^{2}}.  \label{lineofsight}
\end{equation}
Substituting this in the previous expression, we have 
\[
1=\frac{R}{R_{L}}4(m+1)\gamma _{L}^{2}\left[ 1-\mu +\frac{1}{4(m+1)\gamma
_{L}^{2}}\left( \frac{R}{R_{L}}\right) ^{m}\right] .
\]
Which can be rewritten as 
\[
1-\mu =\frac{1}{4(m+1)\gamma _{L}^{2}}\left[ \left( \frac{R}{R_{L}}\right)
^{-1}-\left( \frac{R}{R_{L}}\right) ^{m}\right] .
\]
The perpendicular size, in which we are interested, is given by 
\[
R_{\perp }=R\sin \theta =R\sqrt{1-\mu ^{2}}\cong \sqrt{2}R\sqrt{1-\mu }.
\]
It is therefore maximal when $R^{2}(1-\mu )$ is maximal.  
\[
R^{2}(1-\mu )\propto \left( \frac{R}{R_{L}}\right) -\left( \frac{R}{R_{L}}%
\right) ^{m+2},
\]
and the maximum is obtained for 
\[
\frac{R}{R_{L}}=\left( \frac{1}{m+2}\right) ^{1/(m+1)}.
\]
The maximal perpendicular size from which radiation arrives is therefore
given by 
\begin{eqnarray*}
R_{\perp } &=&\sqrt{2}R\sqrt{1-\mu }=\sqrt{2}\frac{R_{L}}{\gamma _{L}}\left( 
\frac{R}{R_{L}}\right) \sqrt{\frac{1}{4(m+1)}\left[ \left( \frac{R}{R_{L}}%
\right) ^{-1}-\left( \frac{R}{R_{L}}\right) ^{m}\right] }= \\
&&\frac{R_{L}/\gamma _{L}}{\sqrt{2(m+2)}}\left( \frac{1}{m+2}\right)
^{1/2(m+1)}.
\end{eqnarray*}
We now only need to get the coefficient for the relation between
$R_{L}$ and $T$ to get a closed expression. Roughly this is given by
$E\sim R^{3}\rho c^{2}\gamma ^{2}$ where $\rho $ is the density just
in front of the shock. The more exact result is given by Blandford
and Mckee (1976; their Eq. 69)\nocite{bm76} as
\begin{equation}
E=\frac{16\pi }{5+4m}R^{3}\rho c^{2}\gamma ^{2}.
\end{equation}
If we write $\rho =AR^{m-3}$, this, together with equation \ref{lineofsight},
translates to 
\[
E=\frac{16\pi A}{5+4m}R_{L}^{m}c^{2}\frac{R_{L}}{4(m+1)cT}\Longrightarrow
R_{L}=\left[ \frac{(5+4m)(m+1)TE}{4\pi cA}\right] ^{1/(m+1)},
\]
\[
R_{L}/\gamma _{L}=2\sqrt{(m+1)cT}\left[ \frac{(5+4m)(m+1)TE}{4\pi cA}\right]
^{1/2(m+1)}.
\]
Therefore:
\[
R_{\perp ,\max }=\sqrt{\frac{2(m+1)cT}{m+2}}\left[ \frac{(5+4m)(m+1)TE}{4\pi
cA(m+2)}\right] ^{1/2(m+1)}.
\]
For a wind we have $m=1$ so 
\[
R_{\perp ,\max ,{\rm WIND}}=\sqrt{\frac{4cT}{3}}\left[ \frac{3TE}{2\pi
cA}\right] ^{1/4}=1.1\times 10^{17}{\rm cm}\left(
\frac{1+z}{2}\right)^{-3/4} E_{52}^{1/4}A_{\star
}^{-1/4}(T/15{\rm\, days})^{3/4},
\]
where $E = 10^{52}E_{52}$, we have corrected for redshift $z$ and have
used $A = 5 \times 10^{11} A_{\star}$ g cm$^{-1}$ as in Chevalier and
Li (2000)\nocite{cl00}.  Or in angular size
\[
\theta _{\rm WIND}=2.2\mu{\rm as}\left( \frac{1+z}{2}\right)^{-3/4}
D_{A,28}^{-1}E_{52}^{1/4}A_{\star }^{-1/4}(T/15{\rm\,days})^{3/4},
\]
where $D_{A,28}$ is the angular distance in units of $10^{28}$ cm.
For ISM models we have $m=3$, and
\[
\theta _{\rm ISM}=2.8\mu{\rm as}\left( \frac{1+z}{2}\right)^{-5/8}
D_{A,28}^{-1}E_{52}^{1/8}n_1^{-1/8}(T/15{\rm\,days})^{5/8},
\]
where $n_1$ is the density of the ISM in particles cm$^{-3}$.  We can
now obtain an estimate of the angular diameter for a jet as
follows. During sideways expansion of the jet, the radius R of the
fireball is practically constant, $\gamma \propto t^{-1/2}$
(\cite{sph99}) and so $R_{\perp ,\max ,{\rm JET}} \propto t^{1/2}$. At the
time of the jet break $t_{\rm JET}$ we thus roughly have $R_{\perp
,\max,{\rm JET}} = R_{\perp ,\max ,{\rm ISM}}(t_{\rm JET})\times (t/t_{\rm
JET})^{1/2}$, i.e.
\[
\theta _{\rm JET}=1.7\mu{\rm as}\left( \frac{1+z}{2}\right)^{-5/8}
D_{A,28}^{-1}E_{52}^{1/8}n_1^{-1/8}(t_{\rm
JET}/8{\rm\,hr})^{1/8}(T/15{\rm\,days})^{1/2}.
\]

\end{document}